\documentclass[12pt,a4,reqno]{amsart}
\usepackage{geometry}                
\usepackage{epstopdf}
\usepackage{bbm}
\usepackage{amsmath}
\usepackage{amsfonts}
\usepackage{amssymb}
\usepackage{latexsym}
\usepackage{graphicx}
\usepackage[latin1]{inputenc}
\usepackage[T1]{fontenc}
\usepackage{mathrsfs}
\newcommand{\ssQ}{{\mathrm{\scriptscriptstyle{Q}}}}

\title{Another derivation of the geometrical KPZ relations}
\author{François David}
\address[François David]{Institut de Physique Théorique,
CNRS, URA 2306, F-91191 Gif-sur-Yvette, France
CEA, IPhT, F-91191 Gif-sur-Yvette, France}
\email{francois.david@cea.fr}
\author{Michel Bauer}
\address[Michel Bauer]{Institut de Physique Théorique,
CEA, IPhT, F-91191 Gif-sur-Yvette, France
CNRS, URA 2306, F-91191 Gif-sur-Yvette, France 
and
Laboratoire de Physique Théorique de l'Ecole Normale Supérieure, CNRS, UMR 8549,
24 rue Lhomond, 75231 Paris Cedex, France}
\email{michel.bauer@cea.fr}
\date{October 16, 2008}       

\begin{document}

\begin{abstract}
We give a physicist's derivation of the geometrical (in the spirit of Duplantier-Sheffield) KPZ relations, via heat kernel methods. It gives a covariant way to define neighborhoods of fractals in 2d quantum gravity, and shows that these relations are in the realm of conformal field theory.
\end{abstract}
\maketitle
\begin{center}

\end{center}

\medskip

\label{intro}
The Knizhnik-Polyakov-Zamolodchikov (KPZ) relations relate the conformal weights $\Delta^{\!\scriptscriptstyle{0}}$ of the (primary) fields operators of a two dimensional conformal field theory (CFT) to the scaling dimensions  $\Delta$ of these operators when this theory is coupled to two dimensional quantum gravity. 
They read (for unitary CFT  with central charge $c$, as well as for many interesting non-unitary CFT corresponding to geometrical models)
\begin{equation}
\label{KPZ}
\Delta^{\!\scriptscriptstyle{0}}=\Delta + {\gamma^2\over 4} \Delta(\Delta-1)
\ ,\quad 
\gamma=\sqrt{{25-c\over 6}}-\sqrt{{1-c\over 6}}
\end{equation}
The initial derivation of the KPZ relations was obtained by quantizing 2d gravity in a light cone gauge \cite{KPZ88}. The $\Delta$'s appear as weights for the  $SL(2,\mathbb{R})$ current algebra in the resulting effective theory. 
Another derivation is provided by using the conformal gauge \cite{David88a}, then the effective theory is known to be the celebrated quantum Liouville theory \cite{Pol81}. The scaling dimensions $\Delta$'s are determined by the same anomaly consistency conditions (absence of conformal/gravitational anomalies) for the field operators as the conditions holding for the Liouville theory itself. They have been generalized to supersymmetric theories \cite{DistlerKawai88}.

The $\Delta$'s can be extracted from the scaling behaviour of the correlation functions for the quantum CFT+gravitation theory.
Besides numerous  explicit calculations of the correlation functions in the Liouville theory \cite{Liouville}, 
these KPZ relations have been extensively checked to hold in the continuum limit of discretized models of 2d gravity constructed by random matrices and discrete random surfaces models \cite{Matrix}.
These ``algebraic KPZ  relations'' are thus perfectly sound and mathematically meaningfull.
They are now an important ingredient in the theory on non-critical strings, topological strings, etc.
 
Many interesting conformal field theories can be constructed as the scaling limit of 2d statistical models expressed in tems of random geometrical objects on the plane. This is the case for polymers, interacting random walk models, random loops and O(n) models, percolation custers, interface models, Hamiltonian walks and travelling salesman problems, etc...
The scaling operators can be viewed as creating ``geometrical objects" (domains and clusters boundary, contact points, defect lines, etc...) in those models and the conformal weights $\Delta^{\!\scriptscriptstyle{0}}$ are related to the fractal (and multifractal) dimensions of these geometrical objects.
Often these geometrical statistical models can be constructed on a random lattice, and the $\Delta$'s are associated to the scaling dimensions of the corresponding geometrical objects in this random geometry.
For these models the KPZ relations have thus a geometrical interpretation. 
This correspondance has been used by Duplantier \cite{Dup99} to study the multifractal geometrical features of many 2d models.

Recently Duplantier and Sheffield have introduced a new and very interesting probabilistic approach to these geometric KPZ relations \cite{DS08}. Given a fractal (possibly random) set $X$ in the plane $\mathbb{R}^2$ (with its standard measure $dz$),
compare its standard Hausdorff dimension $d_H=2-2x$  with its ``quantum'' Hausdorff dimension $d^\ssQ_H=2-2\Delta$ in the Liouville random measure $d\mu(z)\propto\mathrm{e}^{\gamma\varphi(z)}dz$ where $\varphi(z)$ is the Gaussian free field (corresponding to the Liouville field). It is proven in \cite{DS08,RV08} that the relation between $\Delta$ and $x$ is exactly the KPZ relation (\ref{KPZ}) between $\Delta^{\!\scriptscriptstyle{0}}$ and $\Delta$.
These ideas have also been used in \cite{BS08,RV08} to prove similar formulas for some one dimensional random multifractal measures, of interest for some other problems. 

Besides their mathematical interest, these results raise very interesting questions about 2d quantum gravity.

\noindent (1)\ This approach relies on a crucial assumption, in common with the CFT treatment of \cite{David88a, DistlerKawai88}, namely that the Liouville theory is the correct effective theory for 2d gravity, with its couplings fixed by the anomaly consistency condition.
The treatment of the effective theory is however very different, since based on rigorous probabilistic methods. 

\noindent (2)\  These geometric KPZ relations have a large but not complete overlap with the original algebraic KPZ relations. On one hand not all CFT have a purely geometric formulation. On the other hand, and more importantly, most fractal sets $X$ in the plane (deterministic fractals, random but not conformally invariant fractals) do not correspond to some CFT observables. It is also not possible in general to construct similar fractals in a random metric background (e.g. on a large discrete random surface).

\noindent (3)\  In \cite{DS08} the quantum Hausdorff dimension of the fractal $X$ is constructed by treating $\mathrm{e}^{\gamma\varphi(z)}$ as a random measure, but not as a random metric. Indeed $d^\ssQ_H$ is estimated by standard methods of ball coverings or square box decompositions in the plane. Thus the Liouville field $\varphi$ defines a random ``quantum'' measure, but the underlying geometry of the plane stays ``classical''.
The reason seems technical, since the problem of defining and studying ``Riemannian balls'' (defined in term of the geodesic distance) in a random metric is a difficult one. 
Thus it is not completely obvious why the probabilistic techniques of \cite{DS08,RV08} give the ``right'' result.

In this note we give a field theoretical derivation of the geometrical KPZ relations, using CFT techniques. 
This derivation has a drawback at the level of mathematical rigor - we work at the level of quantum field theory physics, not at the level of pure mathematics and probability theory - but has some advantages. Besides providing an alternate, simple and short derivation of the results of \cite{DS08} accessible to theoretical physicists, we formulate the problem of the geometric KPZ relations in a covariant way, by defining the quantum Hausdorff dimension with respect to the quantum metric, not only the quantum measure. This is done by using heat kernel techniques. Thus it can be used to define consistently the Hausdorff dimension of a (random) geometrical object  on a discrete random geometry, for instance to test the geometrical KPZ relations in numerical simulations.

\medskip
\label{text}
Let us consider a fractal set $X$ in the complex plane with fractal dimension $d_H=2-2x$ (for simplicity we do not consider the case of multifractal sets). For consistency and in order to have a large distance IR regulator one should for instance consider that $X$ lies in a compact domain $D\subset\mathbb{R}^2$.
The flat measure $d\mu(z)=dz$  on $\mathbb{R}^2$ induces a measure $d\mu_X(z)$ with support $X$, which has dimension $d_H$. This measure is constructed (in a loose sense) for instance by approximating $X$ by some ``fat covering'' $X_\epsilon$ of $X$ by circles of radius $\sim\epsilon$, and defining $d\mu_X$ as the limit of the standard flat measure restricted on $X_\epsilon$, properly rescaled by the factor $\epsilon^{-d_H}$.
This means in particular that if we choose a point $z_0\in X$ and measure the volume of $X$ in the disc of radius $r$ centered at $z_0$, this scales for small $r$ as
\begin{equation}
\label{VXflat}
V_X(z_0,r)=\int_{|z-z_0|\le r} \hskip -2.em d\mu_X(z)\  \mathop{\simeq}_{r\to 0}\  (r^2)^{1-x}
\end{equation}
If we consider the situation of a smooth conformal Riemannian metric $\mathbf{g}(z)=\mathrm{e}^{\gamma\varphi_{0}(z)}\mathbf{1}$ on the plane, with $\varphi_0(z)$ a smooth function ($\gamma>0$ plays no role at that stage), the measure on the plane is now $d\mu^{\varphi_0}(z)=dz\,\mathrm{e}^{\gamma\varphi_0(z)}$ and the induced measure on the fractal $X$ is
\begin{equation}
\label{muXphi0}
d\mu^{\varphi_0}_X(z)=d\mu_X(z)\ \mathrm{e}^{\gamma(1-x)\varphi_0(z)}
\end{equation}
Indeed the measure stays local and we can locally treat the metric as constant.
The local scaling (\ref{VXflat}) is of course still valid.

\medskip
\label{quantum}
Following \cite{DS08} we now consider the quantum Liouville case, where the metric $\mathbf{g}(z)$ is a random variable, still of the form $\mathrm{e}^{\gamma\varphi(z)}$, where $\varphi(z)$ is a random massless free field.
The measure $d\mu^{\varphi}_X(z)$ is now a random measure with support on $X$, and the question is to compute this measure and its ``quantum dimension'' $d^\ssQ_H=2-2\Delta$. 
As shown in \cite{DS08}, and as expected on general grounds, (\ref{muXphi0}) cannot be stays correct and $\Delta$ must be different from $x$. 
Indeed $\varphi$ fluctuates at arbitrarily small distance scales $a\ll\epsilon$ much smaller than the  ``regulator'' $\epsilon$ used to define the measure and its dimension, so that the correct limit $a\to 0$, then $\epsilon\to 0$ differs from the naive one $\epsilon\to 0$, then $a\to 0$ (this is the usual renormalisation phenomenon).
This quantum measure must still be local, and if it has  scaling dimension $\Delta$ it must be of the form
\begin{equation}
\label{muXphi}
d\mu^{\varphi}_X(z)\propto d\mu_X(z)\,\mathrm{e}^{\gamma(1-\Delta)\varphi(z)}
\end{equation}
At that stage this must be considered as an ansatz. 
We shall show that $\Delta$ can be easily calculated by a self-consistency scaling argument.

\medskip
For this argument we must extend to the quantum case the scaling (\ref{VXflat}) for the volume of $X$ in a disk of radius $r$, with the exponent $d_H=2-2x$ replaced by $d^\ssQ_H=2-2\Delta$ in the r.h.s. of (\ref{VXflat}), but we must take a covariant definition of the ``disk of size $r$'' around $z_0$. 
One would like to consider the geodesic disk $B_{z_0,r}=\{z;\, d_{\varphi_0}(z,z_0)\le r\}$ with $d_{\varphi_0}(z,z_0)$ the geodesic distance in the metric $\mathbf{g}$, but this becomes problematic in a random metric. 
Instead we choose to define the neibourhood of $z_0$ as the ``domain'' filled by a diffusion process at time $t=r^2$, i.e. by using the heat kernel in the random metric $\mathbf{g}$.

\medskip
\label{HKclassical}
Let us first consider the case of a classical (non-fluctuationg) smooth metric, i.e. a smooth field $\varphi_0(z)$.
The heat kernel $K^{\varphi_0}(z,z_0;t)$ is the integral kernel for the exponential of the Laplacian
\begin{equation}
\label{KTp0b}
K^{\varphi_0}(z,z';t)=\langle z|\mathrm{e}^{t\Delta^{\varphi_0}}|z'\rangle
\end{equation}
where $\Delta_z^{\varphi_0}$ is the covariant Laplace-Beltrami operator in the metric $\mathbf{g}$
\begin{equation}
\label{Laplp0}
\Delta_z^{\varphi_0}=\mathrm{e}^{-\gamma\varphi_0(z)}\,\Delta_z
\ ,\quad \Delta_z=4{\partial\over\partial z}{\partial\over\partial \bar z}
\end{equation}
The heat kernel $K(z,z_0;t)$ is a scalar function of $z$ and it is concentrated in a region of size $r=\sqrt{t}$ around $z_0$ at short times $t$. It is a standard tool in quantum field theory (in particular to study QFT in general background fields and metrics), in differential geometry and in topology. It has been already considered in the context of 2d gravity \cite{HK2d}.
The heat kernel in flat space ($\varphi_0(z)=0$) is simply
\begin{equation}
\label{HKt0}
K^{0}(z,z';t)={1\over 4\pi t}\ \exp\left({-{|z-z'|^2\over 4t}}\right)
\end{equation}
We choose to extract the short distance behavior of the fractal measure $d\mu^{\varphi_0}_X$ from its convolution with the heat kernel. We thus consider the average integral
\begin{equation}
\label{BXtp0}
B_X^{\varphi_0}(z_0,t)=\int_D d\mu^{\varphi_0}_X(z)\, K^{\varphi_0}_t(z,z_0)
\end{equation}
It is convenient to study the small $t$ behavior of $B_X$ through its Mellin-Barnes transform
\begin{equation}
\label{MXsp0}
M_X^{\varphi_0}(z_0,s)=\int_0^\infty dt\, t^{s-1}\, B_X^{\varphi_0}(z_0,t)=\int_D d\mu^{\varphi_0}_X(d
z)\, M^{\varphi_0}(z,z_0;s)
\end{equation}
with $M^{\varphi_0}(z,z_0;s)$ the Mellin-Barnes transform of the heat kernel $K^{\varphi_0}(z,z_0;t)$
\begin{equation}
\label{Msp0}
M^{\varphi_0}(z,z';s)=\Gamma(s)\,\langle z| \left({1\over -\Delta_z^{\varphi_0}}\right)^s|z'\rangle
\end{equation}
Of course in a smooth metric at short distance $M^{\varphi_0}(z,z';s)$ behaves as in flat space
\begin{equation}
\label{Msz0}
M^{\varphi_0}(z,z';s)\ \mathop{\simeq}_{z\to z_0}\ M^{0}(z-z';s)= \Gamma(s)\,\langle z| \left({1\over -\Delta_z}\right)^s|z'\rangle \simeq |z-z'|^{2s-2}
\end{equation}
The integral (\ref{MXsp0}) defining $M_X^{\varphi_0}(z_0,s)$ behaves at small distance $z\to z_0$ as
\begin{equation}
\label{intms}
\int d\mu_X(z)\ |z-z_0|^{ 2-2s}
\end{equation}
and the short distance behaviour of the fractal measure $d\mu_X$, given by (\ref{VXflat}), implies that the integral (\ref{MXsp0}) is convergent as long as $s>x$, and therefore that the Mellin transform $M_X^{\varphi_0}(z_0,s)$ is analytic as long as $\mathrm{Re}(s)>x$, and has a singularity (a pole) at $s=x$.
By the inverse Mellin transform formula the original function behaves at small $t$ as
\begin{equation}
\label{BXp0t}
B_X^{\varphi_0}(z_0,t)\ \simeq\ t^{-x}\ ,\quad t\to 0
\end{equation}
as expected, and as in the flat space case.

\medskip
\label{HKquantum}
We now consider the quantum case, where $\varphi(z)$ is not a fixed smooth metric, but a random massless free field corresponding to the Liouville model. The action for $\varphi$ is normalized as in \cite{DS08}
\begin{equation}
\label{SLiouv}
S[\varphi]={1\over 4\pi}\int dz\, (\nabla\varphi(z))^2
\end{equation}
so that the propagator (the covariance matrix) is simply (at short distance)
\begin{equation}
\label{prop0}
\langle\varphi(z)\varphi(z')\rangle=G_0(z,z')\simeq-\log |z-z'|
\end{equation}
and the ``coupling constant'' $\gamma$ which enters in the random metric $\mathbf{g}=\mathrm{e}^{\gamma\varphi(z)}\mathbf{1}$ is $0\le\gamma\le 2$. There will be UV divergences in the calculations involving the metric, they will be taken into account by multiplicative renormalisation of the metric and of the measures in the standard way (normal products) and  we shall not need to make them more precise.

As argued above, the measure $d\mu_X^\varphi(z)$ on the fractal $X$ is now also a random measure, locally correlated to $\varphi$, and taken to be of the form (\ref{muXphi}). A priori $\Delta\neq x$ since the dimension of the measure is modified by the short distance fluctuations of the metric.
The quantum average of the fractal measure around $z_0$ is now defined as 
\begin{equation}
\label{BXtp0}
B_X^\ssQ(z_0,t)=\left<\int_D d\mu^{\varphi}_X(z)\, K^{\varphi}(z,z_0;t)\right>_\varphi
\end{equation}
and we shall compute the quantum scaling exponent $\Delta$ for $X$ by noting that $B_X^\ssQ$ must obey  the self-consistent short time scaling
\begin{equation}
\label{BQXscal}
B_X^\ssQ(z_0,t)\simeq t^{-\Delta}\ ,\quad t\to 0
\end{equation}
As previously we consider the Mellin-Barnes transform of $B_X^\ssQ(z_0,t)$, which reads
\begin{equation}
\label{BXtp0}
M_X^\ssQ(z_0,s)=\left<\int_D d\mu^{\varphi}_X(z)\, M^{\varphi}(z,z_0;s)\right>_\varphi
= \Gamma(s) \int_D d\mu_X(z)\, \left<\mathrm{e}^{\gamma(1-\Delta)\varphi(z)}\,M^{\varphi}(z,z_0;s)\right>_\varphi
\end{equation}
The singularity in the $s$ variable still comes from the short distance behavior of the integrand. 
We claim that
\begin{equation}
\label{vevMexp}
\left<\mathrm{e}^{\gamma(1-\Delta)\varphi(z)}\,M^{\varphi}(z,z_0;s)\right>_\varphi
\propto |z-z_0|^{2s-2+{\gamma^2\over 2}(s-1)(2\Delta-s)}
\end{equation}
To show this we write
\begin{equation}
\label{Ints}
\mathrm{e}^{\gamma(1-\Delta)\varphi(z)}\,M^{\varphi}(z,z_0;s)
=
\mathrm{e}^{\gamma(1-\Delta)\varphi(z)} \langle z|\left({1\over -\Delta_z^\varphi}\right)^s|z_0\rangle
\end{equation}
and we use the usual ``replica'' trick. We study (\ref{Ints}) for positive integers $s$ and we analytically continue the result to the interesting domain $0<s<1$.
For $s$ integer we use (\ref{Laplp0}) to write the propagator (the inverse of the Laplacian) as $(-\Delta_z^\varphi)^{-1}=(-\Delta_z)^{-1}\mathrm{e}^{\gamma\varphi}$ and to rewrite the r.h.s. of (\ref{Ints}) as
\begin{align}
\label{intvev}
\iint dz_1\cdots dz_{s-1}\ \mathrm{e}^{\gamma(1-\Delta_z)\varphi(z)}
\langle z|\left({1\over -\Delta_z}\right)|z_{s-1}\rangle\,
\mathrm{e}^{\gamma\varphi(z_{s-1})}
\langle z_{s-1}|\left({1\over -\Delta_z}\right)|z_{s-2}\rangle
\cdots\nonumber\\
\cdots
\mathrm{e}^{\gamma\varphi(z_{2})}
\langle z_{2}|\left({1\over -\Delta_z}\right)|z_{1}\rangle
\mathrm{e}^{\gamma\varphi(z_{1})}
\langle z_{1}|\left({1\over -\Delta_z}\right)|z_{0}\rangle
\end{align}
where $\langle z|(-\Delta_z)^{-1}|z'\rangle$ is the massless propagator in flat space.
The quantum $\varphi$ average is performed easily using Wick theorem. At short distances it reads
\begin{equation}
\label{vevexp}
\left<
\mathrm{e}^{\gamma(1-\Delta)\varphi(z)}\mathrm{e}^{\gamma\varphi(z_{s-1})}\cdots\mathrm{e}^{\gamma\varphi(z_{1})}
\right>_\varphi\propto \prod_{j=1}^{s-1}|z-z_{j}|^{-\gamma^2 (1-\Delta)}\prod_{0<i<j<s}(z_i-z_j|^{-\gamma^2}
\end{equation}
We are interested in the singular part in the $z\to z_0$ expansion of (\ref{intvev}) which comes from the sector where 
all the $|z_j-z_0|$ are of the order $|z-z_0|$, since this will give the dominant contribution (after analytic continuation to $0<s<1$). 
the r.h.s. of (\ref{vevexp}) is of dimension (in $z$) ${-\gamma^2}((1-\Delta)(s-1)+(s-1)(s-2)/2)$ and by power counting we obtain (\ref{vevMexp})
(the logarithms coming from the massless propagators do not change this scaling, and might just give a global $\log[z-z_0|$ for integer $s$).

Now comparing (\ref{vevMexp}) to (\ref{Msz0}) and (\ref{BXtp0})  to (\ref{intms}) we see that the first singularity of $M_X^\ssQ(z_0,s)$ occurs at $s_c$ given by
\begin{equation}
\label{ xsc}
2x-2=2s_c-2+{\gamma^2\over 2}(s_c-1)(2\Delta-s_c)
\end{equation}
and the consistency condition $s_c=\Delta$ implies
\begin{equation}
\label{KPZx}
x=\Delta+{\gamma^2\over 4}\Delta(\Delta-1)
\end{equation}
Q.E.D.

\medskip
\label{Boundary}
The same construction and the same argument can be used to derive the geometric boundary KPZ relations considered also in \cite{DS08}.
The Liouville free field $\varphi$ is defined in a simply connected domain $D$ with a smooth boundary $\partial D$, with free boundary conditions (i.e. Neuman b.c.  $\partial_\perp \varphi=0$ on $\partial D$). 
For simplicity we take for $D$ the upper half plane and for $\partial D$ the real axis $\mathbb{R}$.
Now let $X$ be a fractal subset of $\partial D$, with fractal dimension $\tilde d_H=1-\tilde x$.
If $du$ is the standard (one dimensional) mesure on $\partial D$, the induced mesure with support on $X$ is denoted $d\tilde\mu_X(u)$.
If we first consider a smooth non-fluctuating conformal metric $\mathbf{g}(z)=\mathrm{e}^{\gamma\varphi_0(z)}\mathbf{1}$ in $D$ given by a smooth $\varphi_0(z)$, the induced metric on the boundary is $\mathbf{h}(u)=\mathrm{e}^{\gamma\varphi_0(z)}\mathbf{1}$ and the induced measures on the boundary $\partial D$ and the fractal $X$ are respectively $d\tilde\mu^{\varphi_0}(u)=du\, \mathrm{e}^{\gamma\varphi_0(u)/2}$ and
\begin{equation}
\label{ mesXp0b}
d\tilde\mu^{\varphi_0}_X(u)=d\tilde\mu_X(u)\  \mathrm{e}^{{\gamma\over 2}(1-\tilde x)\varphi_0(u)}
\end{equation}
To define the boundary fractal dimension $\tilde d_X$ of $X$ in a covariant way we use the \emph{boundary heat kernel}
$\tilde K^{\varphi_0}(u,u';t)$, solution of the one dimensional diffusion equation \emph{on the boundary} $\partial D$
\begin{equation}
\label{HKp0b}
\tilde K^{\varphi_0}(u,u';t)=\langle u|\mathrm{e}^{t\,\tilde\Delta_u^{\varphi_0}}|u'\rangle
\end{equation}
where $\tilde\Delta_u^{\varphi_0}$ is the one dimensional Laplace-Beltrami operator on $\partial D$ in the metric $\mathbf{h}(u)$
\begin{equation}
\label{LPbD2b}
\tilde\Delta_u^{\varphi_0}=
\mathrm{e}^{-{\gamma\over 2}\varphi_0(u)}\ \partial_u \ \mathrm{e}^{-{\gamma\over 2}\varphi_0(u)}\ \partial_u\ 
=
\left(D_u^{\varphi_0}\right)^2\ ,\quad D_u^{\varphi_0}=\mathrm{e}^{-{\gamma\over 2}\varphi_0(u)}\partial_u
\end{equation}
In the flat metric $\varphi_0=0$ it is of course simply
\begin{equation}
\label{HKt0}
\tilde K^{0}(u,u';t)={1\over \sqrt{4\pi\,t}}\,\exp\left({-{|u-u'|^2\over 4t}}\right)
\end{equation}
The average of the boundary heat kernel over the boundary fractal $X$ scales at small time as
\begin{equation}
\label{BXp0b}
\tilde B_X^{\varphi_0}(u_0,t)=\int_{\partial D} d\tilde\mu^{\varphi_0}_X(u)\  \tilde K^{\varphi_0}(u,u_0;t)\ \simeq\ t^{-\tilde x/2}
\ ,\quad t\to 0
\end{equation}
Equivalently its Mellin-Barnes transform 
\begin{equation}
\label{ MBXp0b}
\tilde M_X^{\varphi_0}(u_0,s)= \int_{\partial D} d\tilde\mu^{\varphi_0}_X(u)
\ \tilde M^{\varphi_0}(u,u_0;s)
\ ,\quad
\ \tilde M^{\varphi_0}(u,u_0;s)=\Gamma(s)\ 
\langle u | \left({1\over -D^{\varphi_0}_u}\right)^{2s} | u_0 \rangle
\end{equation}
has its first pole at $s=\tilde x/2$.

In the quantum case, the fractal dimension is renormalized as $\tilde d^\ssQ_H=(1-\tilde\Delta)$ and the boundary measure on $X$ is taken to be
\begin{equation}
\label{mesQb}
d\tilde\mu^\ssQ_X(u)=d\tilde\mu_X(u)\  \mathrm{e}^{{\gamma\over 2}(1-\tilde \Delta)\varphi(u)}
\end{equation}
$\tilde\Delta$ is fixed by the self-consistency condition for the small time scaling for the boundary heat kernel average
\begin{equation}
\label{MBXb}
\tilde B_X^\ssQ(u_0,t)=\int_{\partial D} {\langle{d\tilde\mu^{\varphi}_X(u) \tilde K^{\varphi}(u,u_0;t)}\rangle}_\varphi\ \simeq\ t^{-\tilde \Delta/2}
\ ,\quad t\to 0
\end{equation}
or equivalently that its Mellin-Barnes transform $\tilde M_X^\ssQ(u_0,s)$ has its first pole at $s_c=\tilde\Delta/2$.
We thus have  compute the short distance behavior of the v.e.v. of  the measure $d\tilde\mu^\ssQ_X$ times the Mellin-Barnes transform of the boundary heat kernel $\tilde M^{\varphi}(u,u_0;s)$
The calculation goes along the same lines as in the bulk case. But now the e.v. of exponentials of $\varphi$ are taken on the boundary.
The Neuman boundary conditions implies that the short distance behavior of the correlator is now
\begin{equation}
\label{propb}
\langle\varphi(u)\varphi(u')\rangle=\tilde G_0(u,u')\simeq-2\,\log|u-u'|\quad,
\end{equation}
while $\tilde M^{\varphi}(u,u_0;s)$ is the kernel for the boundary operator $(-D_u)^{-2s}$ instead of the bulk operator $(-\Delta_z)^{-s}$.
The final result is
\begin{equation}
\label{vevscb}
{\langle
\mathrm{e}^{{\gamma\over 2}(1-\tilde \Delta)\varphi(u)}
\ \tilde M^{\varphi}(u,u_0;s)
\rangle}_\varphi
\ \mathop{\propto}_{u\to u_0}\ 
|u-u_0|^{2s-1+{\gamma^2\over 2}(2s-1)(\tilde\Delta-s))}
\end{equation}
This implies that $\tilde M^{\varphi}(u,u_0;s)$ has its first singularity at $s_c$ given by
\begin{equation}
\label{xscb}
\tilde x-1=2s_c-1+{\gamma^2\over 2}(2s_c-1)(\tilde\Delta-s_c))
\end{equation}
and the consistency condition $s_c=\tilde\Delta/2$ implies the boundary KPZ relation
\begin{equation}
\label{KPZb}
\tilde x=\tilde\Delta +{\gamma^2\over 4}\tilde\Delta(\tilde\Delta-1)
\end{equation}
similar to (\ref{KPZx}).

\medskip
\label{Discussion}
Let us discuss our results.
Formula (\ref{vevMexp}) (and its boundary counterpart (\ref{vevscb})) is the crux of the argument. 
It is obtained here by a replica argument. Since the heat kernel is the solution of a diffusion equation, it can be studied by probabilistic methods, and these methods could probably be used to obtain a more rigorous derivation of (\ref{vevMexp}).

The fact that the heat kernel is a natural object to formulate in a covariant way the geometric KPZ relations is not surprising. The heat kernel has simple properties under conformal transformations. In particular its short distance and time behaviors are related to the spectral dimension of space, and it is known that in 2d quantum gravity the spectral dimension of space-time is still $d_s=2$ (and the spectral dimension of its boundary $\tilde d_s=1$) \cite{HK2d}.
We expect the situation to be quite different and interesting to study when dealing with the intrinsic quantum Hausdorff dimension (defined in term of the geodesic distance), which is know to be $d_H^\ssQ=4$ in the $c=0$ ($\gamma=8/3$) case \cite{KKMW93}, but is very difficult to study in the general case \cite{David92}.


\begin{thebibliography}{99}
\bibitem{KPZ88} V. G. Knizhnik, A. M. Polyakov \& A. B. Zamolodchikov, \textsl{Fractal structure of 2D-quantum gravity}, Modern Phys. Lett. A, 3 (1988) 819-826.

\bibitem{Pol81} A. M. Polyakov, \textsl{Quantum geometry of bosonic strings}, Phys. Lett. B, 103 (1981) 207-210, \\
 A. M. Polyakov, \textsl{Quantum geometry of fermionic strings}, Phys. Lett. B, 103 (1981)211-213.
 
\bibitem{David88a} F. David, \textsl{Sur l'entropie des surfaces aléatoires}, C. R. Acad. Sci. Paris, 307, II (1988) 1051-1053, \\
F. David, \textsl{Conformal field theories coupled to 2-d gravity in the conformal gauge}, Mod. Phys. Lett. A, 3 (1988) 1651-1656.

\bibitem{DistlerKawai88} J. Distler \& H. Kawai, \textsl{Conformal Field Theory and 2D quantum gravity or who's afraid of Joseph Liouville?}, Nucl. Phys. B, 321 (1989) 509-527. 
J. Distler, Z. Hlousek \& H. Kawai, \textsl{Hausdorff Dimension of Continuous Polyakov's Random Surfaces or who' afraid of Joseph Liouville? Part 2},  Int.J.Mod.Phys.A5:1093,1990.

\bibitem{Liouville}
There are numerous reviews on Liouville theory, see for instance:\\
J. Teschner, \textsl{Liouville theory revisited}, Class. Quant. Grav. 18 (2001) R153-R222, arXiv:hep-th/0104158v3,\\
Y. Nakayama, \textsl{Liouville Field Theory ? A decade after the revolution}, Int. J. Mod. Phys. A19 (2004) 2771-2930, arXiv:hep-th/0402009v7.

\bibitem{Matrix}There are also numerous reviews on matrix models and 2d gravity, see for instance:\\
P. Di Francesco, \textsl{2D Quantum Gravity, Matrix Models and Graph Combinatorics}, arXiv:math-ph/0406013,\\
B. Eynard, \textsl{Large N asymptotics of orthogonal polynomials, from integrability to algebraic geometry},  arXiv:math-ph/0503052,\\
I. Kostov, \textsl{Matrix Models as Conformal Field Theories, }\\
in \textsl{ Applications of Random Matrices in Physics}, 
Eds. Brézin E., Kasakov V.A., Serban D., Wiegmann P.B., Zabrodin A.
Les Houches Summer School 2004,
NATO Science Series II 221 (2006)\\
and references therein.

 \bibitem{Dup99} B. Duplantier, \textsl{Conformally Invariant Fractals and Potential Theory}, Phys. Rev. Lett. 84, 1363-1367 (2000).\\
B. Duplantier, \textsl{Conformal Random Geometry}, arXiv:math-ph/0608053v1, in Les Houches Summer School, Session LXXXIII, 2005, \textsl{Mathematical Statistical Physics}, A. Bovier, F. Dunlop, F. den Hollander, A. van Enter and J. Dalibard, eds., pp. 101-217, Elsevier B. V. (2006).

\bibitem{DS08} B. Duplantier \& S. Sheffield, \textsl{Liouville Quantum Gravity and KPZ}, 2008, arXiV:0808.1560v1 [math:PR] 

\bibitem{BS08} I. Benjamini \& O. Schramm, \textsl{KPZ in one dimensional random geometry of multiplicative cascades}, 2008, 
arXiV:0806.1347v1 [math.PR]

\bibitem{RV08} R. Rhodes \& V. Vargas, \textsl{KPZ formula for log-infinitely dividible multifractal random measures}, 2008, arXiV:0807.1036v2 [math.PR]

\bibitem{HK2d} J. Ambjørn, D. Boulatov, J. L. Nielsen, J. Rolf \& Y. Watabiki, \textsl{The Spectral Dimension of 2D Quantum Gravity},	JHEP 9802 (1998) 010, arXiv:hep-th/9801099v1.

\bibitem{KKMW93} H. Kawai, N. Kawamoto, T. Mogani \& Y. Watabiki, \textsl{Transfer Matrix Formalism for Two-Dimensional Quantum Gravity and Fractal Structures of Space-time}, Phys. Lett. B306 (1993) 19-26, arXiv:hep-th/9302133.

\bibitem{David92} F. David, \textsl{What is the intrinsic geometry of two-dimensional quantum gravity?}, Nucl. Phys. B 368 (1992) 671-700.

\end{thebibliography}
\end{document}